\documentclass[fleqn,usenatbib]{mnras}
\usepackage{cite}
\usepackage{pgfplots}
\pgfplotsset{compat=1.14}

\usepackage{newtxtext,newtxmath}

\usepackage[T1]{fontenc}
\usepackage{ae,aecompl}
\usepackage{verbatim}

\usepackage{graphicx}	
\usepackage{amsmath}	

\usepackage{amssymb}	
\usepackage{multicol}
\usepackage{multirow}
\usepackage{pdflscape}	
\usepackage{amsmath} 
\newcommand{\angstrom}{\text{\normalfont\AA}}
\usepackage{color}

\usepackage{times}
\usepackage[space]{grffile}
\usepackage{cleveref}
\usepackage[normalem]{ulem}
\usepackage{natbib}
\usepackage{longtable}
\usepackage{hyperref}
\usepackage{comment}
\bibpunct{(}{)}{;}{a}{}{,}
\usepackage{pdflscape}	
\usepackage{url}
\texorpdfstring

\newcommand{\alphaox}{$\alpha_\mathrm{ox}$}

\newcommand{\xmm}{{\em XMM-Newton}}
\newcommand{\nustar}{{\em NuSTAR }}
\newcommand{\chandra}{{\em Chandra}}
\newcommand{\suzaku}{{\em Suzaku}}

\newcommand{\swift}{{\small \it Swift}}

\newcommand{\xrt}{{\small {\it Swift}/XRT}}
\newcommand{\uvot}{{\small {\it Swift}/UVOT}}

\graphicspath{{./}{pic/}}

\title[Mrk1018]{Long-term and multi-wavelength evolution of a changing-look AGN Mrk 1018}

\author[Bing Lyu et al.]{
Bing Lyu,$^{1,2}$
Zhen Yan,$^{2}$
Wenfei Yu,$^{2}$
Qingwen Wu,$^{1}$\thanks{E-mail:qwwu@hust.edu.cn}
\\
$^{1}$Huazhong University of Science and Technology, School of Physics, 1037 Luoyu Road, Wuhan, 430074, China\\
$^{2}$Shanghai Astronomical Observatory, CAS, 
Nandan Road 80, Shanghai, 200030, China\\
}

\date{Accepted XXX. Received YYY; in original form ZZZ}

\pubyear{2021}

\begin{document}
\label{firstpage}
\pagerange{\pageref{firstpage}--\pageref{lastpage}}
\maketitle

\def\sectionautorefname{Section}
\def\subsectionautorefname{Section}


\begin{abstract}
The physical mechanism for triggering the changing-look phenomenon in active galactic nuclei (AGNs) is still unclear. We explore this issue based on the multi-wavelength spectral and flux variations for a changing-look AGN Mrk~1018 with long-term observations in the X-ray, optical/ultraviolet(UV), and radio bands.  
Both the optical and the X-ray emission experience rapid decay in changing-look phase during 2010--2015, where a re-flare appears in the optical/UV and X-ray bands. We find a time lag of $\sim 20 $ days of optical/UV behind X-ray variations in type 1.9 phase. The 5 GHz radio flux decreases by $\sim 20$\% in type 1.9 phase during 2016--2017. We find both X-ray photon index ($\Gamma$) and the optical-to-X-ray spectral index (\alphaox\,) are anti-correlated with the Eddington scaled 2--10~keV X-ray luminosity ($L_\mathrm{X}/L_\mathrm{Edd}$) in the type 1.9 phase. However, the type 1 phase deviates from these two anti-correlations, which suggests that the change of broad emission lines might be regulated by the evolution of accretion disk (e.g., disappearing of the inner cold disk in the type 1.9 phase). 
\end{abstract}


\section{Introduction}\label{sec:intro}

Type 1 and type 2 active galactic nuclei (AGNs) are classified based on the widths of optical spectral emission lines. Type 1 AGNs show both broad lines ($>$ 1000 $ \rm{km}\, \rm{s}^{-1}$) and narrow lines ($<$ 1000 $ \rm{km}\, \rm{s}^{-1}$), while type 2 AGNs show only narrow lines. In the AGN unification model \citep[e.g.][]{1993ARA&A..31..473A}, type 1 AGNs are viewed face-on with the broad-line region (BLR) visible to the observer, while type 2 AGNs are viewed edge-on with the broad-line region blocked by a putative dusty torus. The sub-classes (e.g., type 1.5, 1.8, and 1.9) are also introduced \citep[see ][]{1976MNRAS.176P..61O,1981ApJ...249..462O} based on the emission-line width and relative strength of the broad-line to the narrow-line. There are broad H$\alpha$ line and very weak or undetectable broad H$\beta$ line in type 1.9, broad H$\alpha$ line and weak broad H$\beta$ line in type 1.8 \citep[see][]{1986ApJ...311..135C}, and comparable H$\alpha$ and H$\beta$ lines in type 1.5.  

In recent years, several tens of so-called changing-look AGNs (CL-AGNs hereafter) have been discovered, which show disappearance/appearance of broad emission lines within a timescale of decades or years \citep[e.g.][]{2014ApJ...796..134D,2014ApJ...788...48S,2015ApJ...800..144L,2016A&A...593L...8M,2016MNRAS.461.1927P,2016ApJ...826..188R,2018ApJ...862..109Y,2019MNRAS.486..123R,2020MNRAS.491.4925G,2020ApJ...890L..29A,2020ApJ...901....1W,2020A&A...638A..91K}, or even months \citep[e.g.][]{2019MNRAS.487.4057K,2019ApJ...883...94T}. The term ``changing-look'' was originally used to describe the changes of AGNs from Compton-thick to Compton-thin (or vice versa) based on the X-ray observations \citep[e.g.][]{2003MNRAS.342..422M}. In this paper, the term ``changing-look" refers to the change in optical emission lines. The physical mechanism of the changing-look phenomena is still under debate. On the one hand, the ``changing-look'' can be attributed to variable obscuration, such as obscuring material moving in or out from our line of sight \citep[e.g.][]{2013MNRAS.436.1615M,2014MNRAS.443.2862A,2015ApJ...815...55R,2018MNRAS.481.2470T}, where the intrinsic emission is roughly unchanged. On the other hand, the changing-look may associate with the variation of intrinsic radiation \citep[e.g., the change of accretion disk][]{1984MNRAS.211P..33P,2014MNRAS.438.3340E}. Correlated variability in the X-ray \citep[e.g.][]{2016MNRAS.461.1927P,2019MNRAS.483L..88P,2020ApJ...898L...1R}, the optical/UV \citep[e.g.][]{2019ApJ...885...44D} and the infrared band \citep[e.g.][]{2017ApJ...846L...7S,2018ApJ...864...27S} with the AGN type change supports variable accretion as the physical origin in some CL-AGNs, which is also supported by that the absorption is roughly unchanged in some CL-AGNs \citep[e.g.][]{2016A&A...593L...9H,2020ApJ...890L..29A,2020ApJ...901....1W}.

The Seyfert galaxy Mrk~1018 at $z=0.042$ has undergone a full cycle with twice types transitions during the past 40 years. It transited from type 1.9 to type 1 between 1979 and 1984 \citep{1986ApJ...311..135C} and returned to type 1.9 after 30 years \citep[see][]{2016A&A...593L...8M,2016A&A...593L...9H,2017A&A...607L...9K}. The optical spectroscopic observations reveal that Mrk~1018 is a type 1 AGN in 2010 December and a type 1.9 AGN in 2015 January \citep{2016A&A...593L...8M,2018ApJ...861...51K}. The optical spectroscopic observation of Mrk~1018 in 2019 October shows a faint broad H$\beta$ line component in \citet{2020A&A...644L...5H}, which is similar to the type 1.9 spectrum as reported in \citet[][]{2016A&A...593L...8M}. 

In this work, we explore the possible physical mechanism for the changing look by performing an extensive data analysis for a CL-AGN of Mrk~1018 in the radio, the optical/UV, and the X-ray bands. We mainly focus on the period of 2005-2019, which is covered by multi-wavelength observations. According to the optical spectroscopic results, we identify that the period of 2005--2010 is the type 1 AGN phase, and 2015--2019 is the type 1.9 AGN phase. The paper is organized as follows. In \autoref{sec:data}, we describe the observations and the data reduction in different wavebands. In \autoref{sec:result}, we present the multi-wavelength observational results. In \autoref{sec:discussion}, we discuss possible physics behind the observational results. Finally, we summarize our results in \autoref{sec:conclusion}. Throughout this work, we use a flat $\Lambda-$CDM cosmological model with $\Omega_{\rm{M}}$=0.27, $\Omega_\Lambda=$0.73 and a Hubble constant of 70 km s$^{-1}$ Mpc$^{-1}$. We adopt the luminosity distance $d_{\rm{L}}=176 $ Mpc and the black hole (BH) mass measurement $\log(M_{\rm{BH}}/M_{\odot})=7.84$ \citep{2017MNRAS.472.3492E,2018MNRAS.480.3898N} for Mrk~1018.

\section{Data reduction and analysis}\label{sec:data}
\subsection{X-ray data analysis}
We analyse the public archival data of \swift, \xmm, \chandra ~and \nustar during the period between 2005 and 2019. We use the cosmic abundances of \citet{2000ApJ...542..914W} and the photoelectric absorption cross sections from \citet{1995A&AS..109..125V}. All the X-ray spectra are fitted by an absorbed power-law model  \texttt{tbabs*zpowerlw} with the absorption by the Galactic hydrogen fixed at $N_{\rm{HI,Gal}}=2.43 \,\times \,10^{20}\, \rm{cm}^{-2}$ \citep[][]{2005A&A...440..775K} since no intrinsic absorption beyond Galactic is detected \citep[see][]{2016A&A...593L...9H,2017A&A...607L...9K}. The 2--10~keV flux are calculated by {\it cflux} component within {\tt XSPEC} (v12.10). The observation information and best-fitting parameters including photon index ($\Gamma$), unabsorbed flux in 2--10~keV ($F_{\rm{2-10~keV}}$) are listed in \autoref{tab:tablexray}. The long-term X-ray light curve in 2--10~keV band is shown in the top panel of \autoref{fig:multi-lc-secondaxis}.

\subsubsection{\xrt\,}
\label{data-xrt}
The X-ray telescope (XRT) on board \swift\, has the highest cadence monitoring observations of Mrk~1018 in the X-ray band, especially after 2015. We reprocess the archive data of \xrt\, observations performed in photon counting mode with {\scriptsize XRTPIPELINE}. The source region is a circle centered at the nucleus of Mrk 1018, the radius of which is determined by the count rate of each observation according to \citet{2009MNRAS.397.1177E}. We use {\scriptsize XSELECT} to extract the source and background spectra. The spectra are grouped by the a minimum of one count per bin. The XRT spectra of Mrk~1018 in the 0.5--10~keV range are fitted using the Bayesian X-ray Analysis software BXA \footnote{\url{http://johannesbuchner.github.io/BXA/index.html}} \citep[][]{2014A&A...564A.125B} which connects the nested sampling algorithm UltraNest \citep{2021JOSS....6.3001B}.  

\subsubsection{\chandra/ACIS-S}
We extract the ACIS-S spectra with CIAO (v4.12) and {\tt CALDB} (v4.9.1). For the observation in November of 2010 (ObsID 12868), which is affected by the pile-up effect, we adopt the fitting results from \citet{2016A&A...593L...9H} which excluded the bright pixels and corrected the photon loss for this observation. The other observations are extracted from a 3$\arcsec$ radius circle and the background spectra are extracted from an annulus with 5$\arcsec$ inner radius and 15$\arcsec$ outer radius \citep[see also][]{2017ApJ...840...11L}. Then the spectra are grouped by a minimum of 20 counts per bin and fitted in the 0.5--8~keV range.

\subsubsection{$XMM-Newton$/EPIC-PN }
We analyse the archival data sets of Mrk~1018 derived by the EPIC-PN on board \xmm\,. The source is observed in 2005 (ObsID 201090201) and 2008 (ObsID 554920301). We reduce the PN data with {\scriptsize EPPROC} in \texttt{SAS-16.1.0}. The source and background regions are a 40$''$ and 60$''$ radius circle, respectively. Each spectrum is grouped by a minimum of 30 counts per bin and fitted in the 2--10~keV range.

\subsubsection{$NuSTAR$}
We analyse the archival data sets of Mrk~1018 on board \nustar, which are reduced through the {\scriptsize NUPIPELINE} task of the {\scriptsize NUSTARDAS} package. The source region is a 50$''$ radius circle at the center of the source, and the background is extracted from the blank region. Each spectrum is grouped by a minimum of 30 counts per bin and fitted in the 3--79~keV range.

\subsection{\uvot\,data analysis}
\label{sec:uvot}
There are six filters in the optical/UV band of \uvot, which are V, B, U, UVW1, UVM2, and UVW2 bands. We use the tool \textit{uvotsource} to do the aperture photometry for each filter of all the observations. The source aperture radius is 5$\arcsec$ and the background is chosen in a blank region with a much larger radius. According to the results in \citet{2018MNRAS.480.3898N}, the emission of the host galaxy is dominated in the V and the B band, we then discard all the results of the V and the B band. 

In order to correct the Galactic extinction, we adopt $E(B-V) = 0.036$ \citep[see][]{2018MNRAS.480.3898N} and $R_{V}=3.1$ for the Galactic extinction and calculate the values of $A_{\lambda}$ for U, UVW1, UVM2, and UVW2 band are  0.18, 0.25, 0.35 and 0.31 assuming the extinction model of \citet{2007ApJ...663..320F}. In order to estimate the intrinsic optical/UV flux from the nucleus, we subtract the contribution from the host galaxy, which is estimated from the broadband spectral modeling result in \citet{2018MNRAS.480.3898N}. The results of the long-term optical and UV light curves from \uvot\, are shown in the middle panel of \autoref{fig:multi-lc-secondaxis} and listed in \autoref{tab:tableuvot}.

\subsection{Radio data analysis}
\label{subsec:vla}
We analyse the archival data of Very Large Array (VLA) observations for Mrk~1018 with \textsc{casa} version 5.3.0 \citep{2007ASPC..376..127M}. For the reduction of the pre-EVLA upgrade VLA data (VLA, project ID: AU0020, AB0476, AB0540, and AB0878), we manually flag and calibrate the data, then clean the image following the instruction\footnote{\url{https://casaguides.nrao.edu/index.php/VLA_5_GHz_continuum_survey_of_Seyfert_galaxies}}. For the new Karl G. Jansky Very Large Array data (JVLA, project ID: 16A-444, 16B-084, and 18B-245), calibrations are performed using script {EVLA\_pipeline1.4.2}\footnote{\url{https://science.nrao.edu/facilities/vla/data-processing/pipeline/scripted-pipeline}}. Different bands are split into different MS files after checking the radio frequency interference and calibration. Then the source is imaged using {\scriptsize TCLEAN} method and integrated flux is estimated via \textsc{imfit} task. The uncertainty of flux density is calculated from $\sigma_\mathrm{S}=\sqrt{(rms)^2+(0.05\times S)^2}$, where $5 \%$ absolute flux error is taken into account, except for the quick look image result in epoch 1 of {\em VLA Sky Survey (VLASS1.1)}, where $15 \%$ system error is considered according to the {\em VLASS Epoch 1 Quick Look Users Guide} \footnote{\url{https://science.nrao.edu/science/surveys/vlass/vlass-epoch-1-quick-look-users-guide}}. The imaging results are listed in Table.~\ref{tab:tableradio}.

To compare between different periods and keep consistent with the radio and X-ray correlation in literature, we convert the radio flux to 5 GHz if it was observed at other wavebands using $S_v \propto v^{-\alpha_\mathrm{R}}$. The radio flux density of Mrk~1018 in the same band varied little before 2015, so we assume the radio spectral index also remains constant during this period. We calculate the $\alpha_\mathrm{L-X} =0.3 \pm 0.08$ using the observations on 50970 MJD (X band) and 52490 MJD (L band), then convert the flux densities at other bands to 5 GHz during this period (see \autoref{tab:tableradio}). There is one observation with two bands (C and X) available on MJD 57481, the $\alpha_\mathrm{C-X} =0.25\pm0.1$ is consistent within uncertainties with previous measured $\alpha_\mathrm{L-X}$. We then use this value to convert the flux densities of other bands to 5 GHz after MJD 57481. The estimated radio light curve at 5 GHz after 2005 is present in the bottom panel of \autoref{fig:multi-lc-secondaxis}.

\section{Results}
\label{sec:result}
\subsection{Multi-wavelength light curves}
\label{sec:multi-lc}
\begin{figure*}
\centering
	\includegraphics[width=0.9\textwidth]{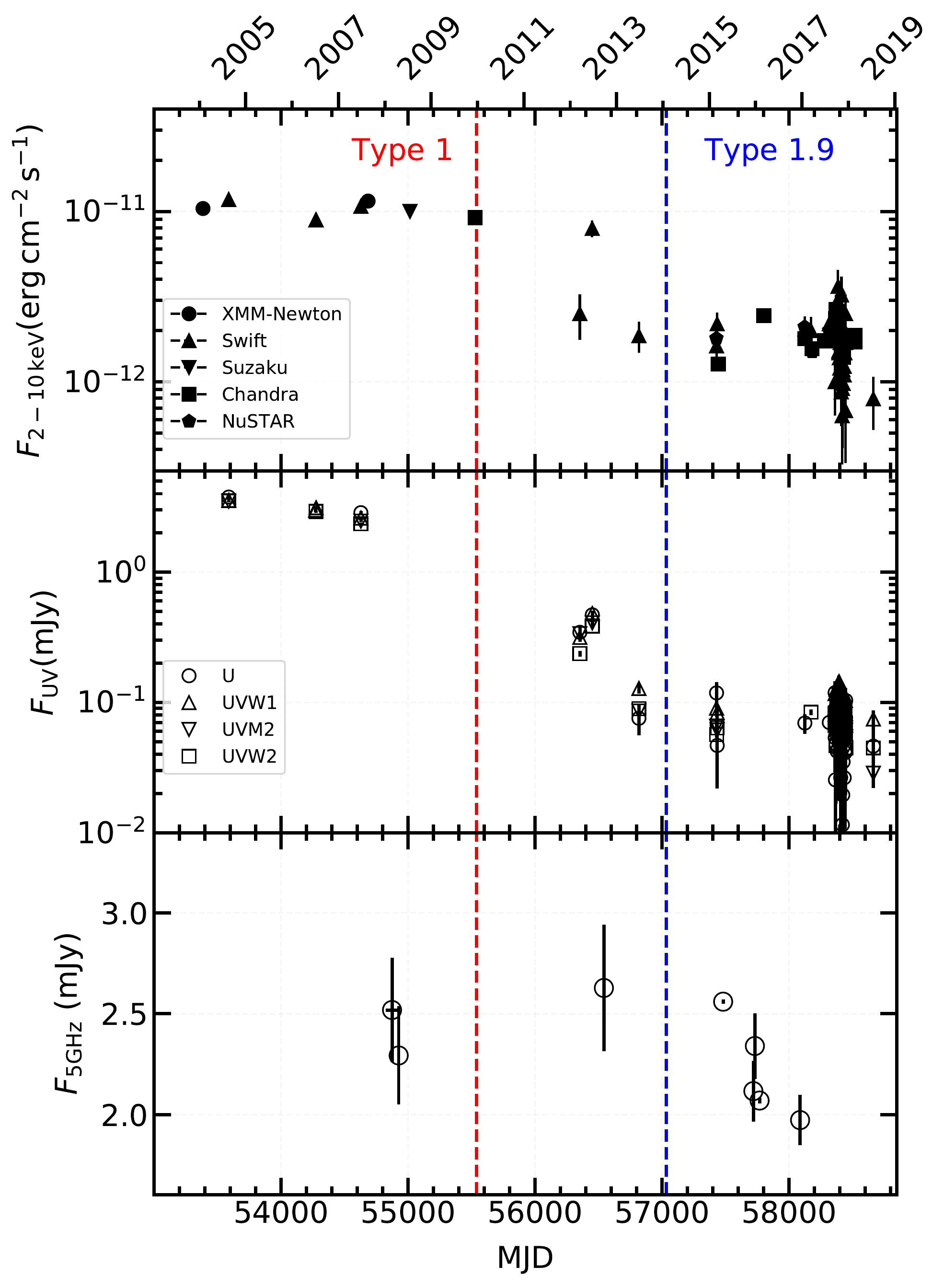}
    \caption{Multi-wavelength light curves of Mrk~1018 between 2005 and 2019. Red and blue vertical dashed lines represent the timeline of optical spectroscopic confirmation at type 1 and type 1.9, respectively. A re-flare during the changing-look phase is found in both the X-ray and the optical/UV bands.}
    \label{fig:multi-lc-secondaxis}
\end{figure*}

Multi-wavelength light curves of Mrk~1018 are presented in \autoref{fig:multi-lc-secondaxis}. Between 2005 and 2010, when Mrk~1018 stayed in the bright type 1 phase, the X-ray flux was roughly unchanged. The optical/UV flux showed a slight decline by a factor $\sim 1.4 $ during 2005--2007. 

Between 2010 and 2015, both the optical/UV and the X-ray flux showed rapid decay \citep[see also ][]{2016A&A...593L...8M,2016A&A...593L...9H}, and Mrk~1018 changed from type 1 into type 1.9. The optical/UV and X-ray flux declined by a factor of $\sim 17$ and $\sim 7.5$, respectively. 

We find a re-flare (around 2013-2014) during the decay phase, while the amplitude of X-ray variation is higher than those in optical/UV bands (see \autoref{fig:multi-lc-secondaxis}, the X-ray flux increase by a factor $\sim 3$ within $\sim$ 100 days then decrease by a factor of $\sim 4.2$. The optical/UV flux increase by a factor of $\sim 1.5$ within $\sim$ 100 days then decrease by a factor of $\sim 3.8$.). After 2015, the source went into the faint type 1.9 phase. The X-ray showed stronger variability ($\sim $ 14 \%) than optical/UV bands ($\sim $ 6 \%) during the type 1.9 phase in 2018.  

The 5 GHz radio flux did not decline during the decay of X-ray and optical/UV emission between 2010 and 2015 and, however, it decreased by $\sim 20$ \% in the type 1.9 phase during 2016--2017 (see \autoref{fig:multi-lc-secondaxis}).

\subsection{Time lag between X-ray and UV variations}
Between 2018 August and 2018 November, \swift\, executed an intensive monitoring campaign on Mrk~1018 (48 visits within 84 days). In order to examine the correlation between X-ray and UV flux variations during this period, we use the interpolation cross correlation function \citep[ICCF;][]{1998PASP..110..660P,2018ascl.soft05032S} with a time lag ($\tau$) range of 0--40 days (around half the overlap). The interpolation time step of 1 day are both applied to X-ray and UV light curves. The flux randomization and random subset selection methods are employed with 10000 realizations in the Monte Carlo simulation to estimate the centroid time lag and the uncertainties \footnote{The code \texttt{pyCCF} is available in \url{http://ascl.net/code/v/1868}}. 

\begin{figure}
\centering
	\includegraphics[width=0.45\textwidth]{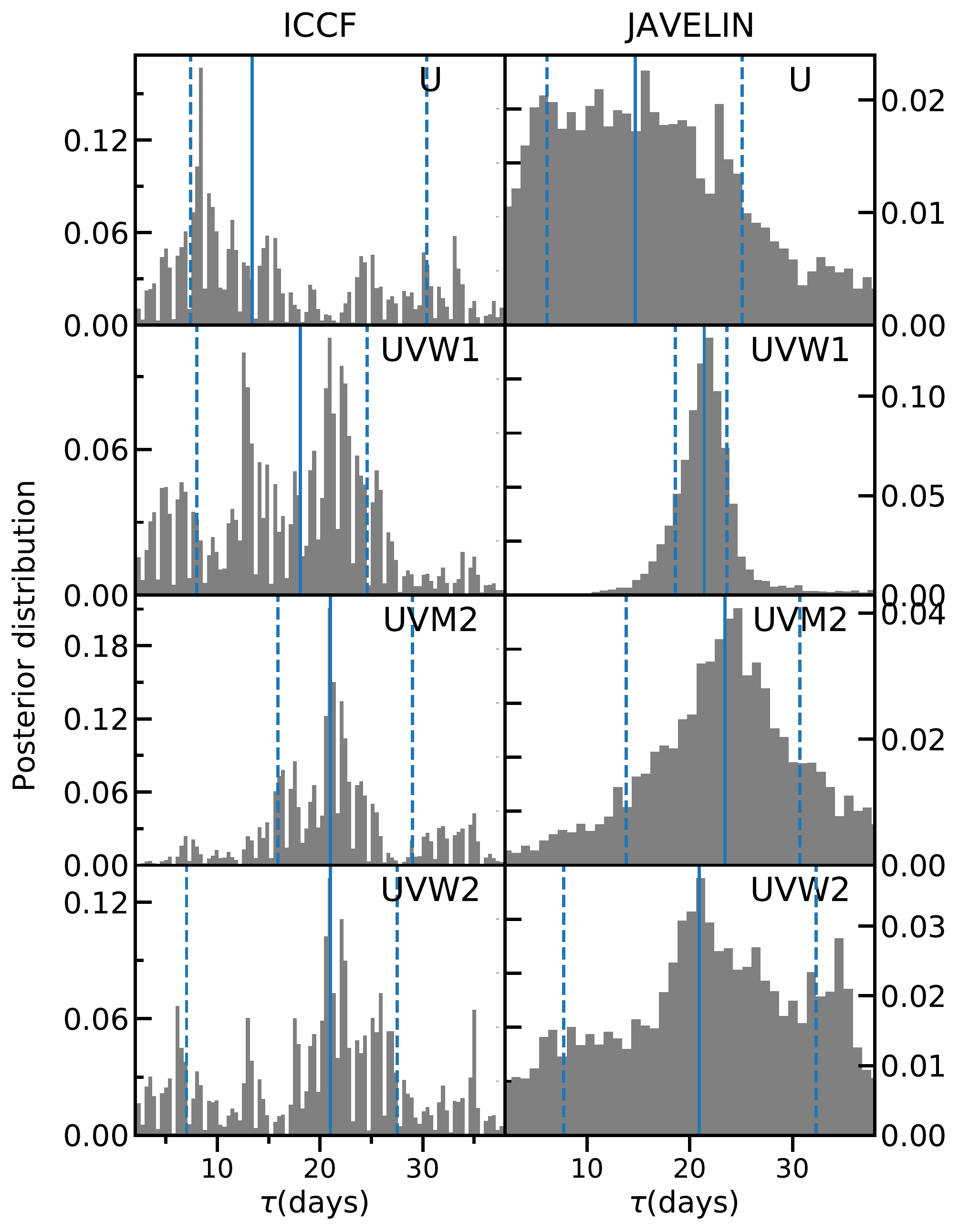}
    \caption{The posterior distribution of $\tau$ with ICCF and \texttt{JAVELIN} methods. The solid lines represent the centroid time lag and the dashed lines correspond to $1\sigma$ range uncertainties.}
    \label{fig:ccf_javelin}
\end{figure}

We also use the \texttt{JAVELIN} algorithm \citep[][]{2011ApJ...735...80Z,2013ApJ...765..106Z} to further examine the time lag that we estimate through the ICCF method. The \texttt{JAVELIN} approach fits the light curves using a damped random walk (DRW) model, convolves them with a top-hat transfer function (TF), and aligns them to recover the time lag and other parameters (such as the amplitude and timescale of the DRW process, the height and width ($w$) of the top-hat transfer function) with the Monte Carlo method. We first restrict the range of time lag $\tau$ and the $w$ to be 0--40 days. The measured time lag $\tau$ from \texttt{JAVELIN} are consistent with those of ICCF method (see \autoref{tab:tablelag}). However, there are two peaks of the posterior distribution of $\tau$ ($\sim$ 20 and 32 days) between X-ray and UVM2 band, which does not agree with the ICCF method. We then restrict the range of $w$ to be 15--40 days and perform an additional simulation. Only one peak at $\sim$ 23 days is shown in the posterior distribution of $\tau$, we then take this as the primary time lag between X-ray and UVM2 bands. So both ICCF and \texttt{JAVELIN} methods give consistent time lags between X-ray and U band at $\sim$14 days, and time lags between X-ray and UVW1/UVM2/UVW2 at $\sim$20 days. The results from the ICCF and \texttt{JAVELIN} methods between X-ray and the four UVOT bands are listed in \autoref{tab:tablelag} and presented in \autoref{fig:ccf_javelin}.


\begin{table}
\renewcommand{\arraystretch}{1.5}
\centering
\caption{{\bf Detected time lag $\tau$ of the UV variations behind the X-ray.} The X-ray band is taken as the reference. The uncertainties refer to $1\sigma$ range.}
\label{tab:tablelag}
\begin{tabular}{lcccccr}
\hline
\hline
 band & method & $\tau$ & correlation coefficient & $p$-value   \\ 
      &        &  [day] &                         &             \\ \hline
U     & ICCF &  $13.4 ^{+17}_{-6} $ & 0.40 & 3.8e-3  \\
UVW1  & ICCF &  $18.1 ^{+6.5}_{-10.1} $ & 0.46 & 1.3e-3 \\
UVM2  & ICCF &  $21.0 ^{+8}_{-5.1} $ & 0.50 & 1.4e-3 \\
UVW2  & ICCF &  $21.0 ^{+6.5}_{-14} $ & 0.37 & 1.7e-2 \\
\hline \hline
band & method & $\tau$ & $w$ range &  $w$  \\ 
     &        &  [day] & [day]  &    [day]  \\ \hline 
U     & \texttt{JAVELIN} &  $ 14.7^{+10.4}_{-8.6} $ & 0--40 & $16.5 ^{+13.4}_{-10.6} $ \\
UVW1  & \texttt{JAVELIN} &  $ 21.4 ^{+2.2}_{-2.8} $  & 0--40 & $9.3 ^{+8.2}_{-5.7} $  \\
UVM2  & \texttt{JAVELIN} &  $ 23.4^{+7.3}_{-9.6} $  & 15--40 &  $ 25.1^{+8.7}_{-7.2} $  \\
UVW2  & \texttt{JAVELIN} &  $ 20.9 ^{+11.4}_{-13.2} $  & 0--40&  $ 12.7 ^{+16.9}_{-10.4} $ \\
\hline \hline
\end{tabular}   
\end{table}


\subsection{$\Gamma$-$L_\mathrm{X}/L_\mathrm{Edd}$ and \alphaox- $L_\mathrm{X}/L_\mathrm{Edd}$ correlation}\label{subsec:xray-uv}
\begin{figure}
\centering
	\includegraphics[width=\linewidth]{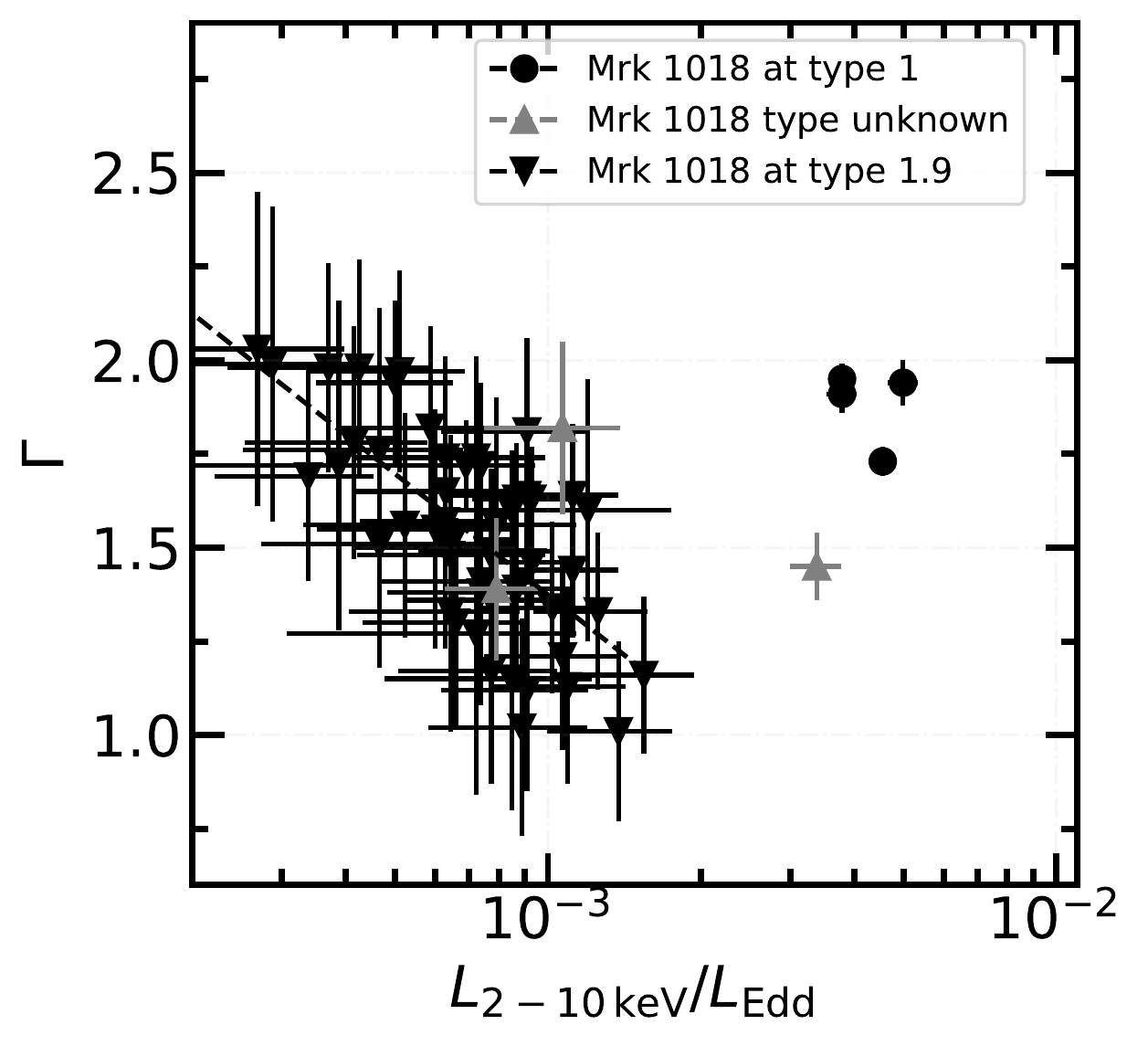}
    \caption{The $\Gamma$ - $L_\mathrm{X}/L_\mathrm{Edd}$ correlation. Only data of \xrt\, are included here. The dashed line represent the best fitting of the negative correlation in the type 1.9 phase.}
    \label{fig:xrayappendgood-Lrateandg-tmap}
\end{figure}

We present the $\Gamma$-$\log{L_\mathrm{X}/L_\mathrm{Edd}}$ correlation of Mrk~1018 in \autoref{fig:xrayappendgood-Lrateandg-tmap}, where only the data of \xrt\, are adopted to avoid the discrepancies of different instruments. We find an evident negative correlation between the photon index and the Eddington-scaled X-ray luminosity in the type 1.9 phase, where the Spearman correlation coefficient is $-0.63$ ($p=5.1\times10^{-7}$). The data in the type 1 phase and the re-flare apparently deviate from the negative correlation (see \autoref{fig:xrayappendgood-Lrateandg-tmap}). 

\begin{figure}
\centering
	\includegraphics[width=\linewidth]{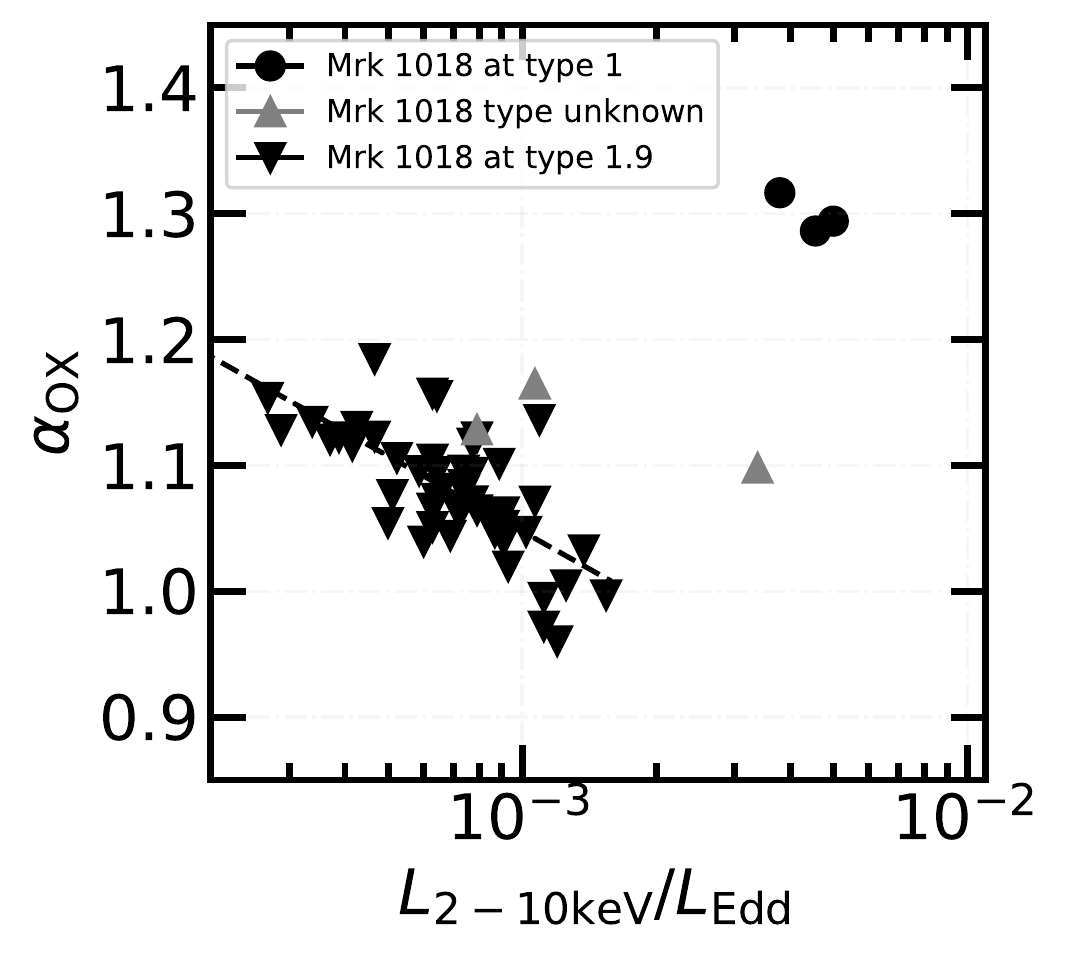}
    \caption{The $\alpha_\mathrm{OX}-L_{\mathrm{X}}/L_\mathrm{Edd}$ correlation. The dashed line represent the best fitting of the negative correlation in the type 1.9 phase.}   
    \label{fig:alpha_ox_lx}
\end{figure}

The correlation of $\alpha_{\rm ox}-\log{L_\mathrm{X}/L_\mathrm{Edd}}$ is also explored in \autoref{fig:alpha_ox_lx}. We calculate the $\alpha_{\rm ox}$ according to the \autoref{definition_alpha_ox}. The $L_\mathrm{UVW1}$ is derived from UVW1 filter of the \uvot\ with central wavelength {2600{$\angstrom$}} and full-width at half max of $\sim 683\angstrom$ \citep{2008MNRAS.383..627P}. The $L_\mathrm{2~ keV}$ is calculated according to \autoref{definition_f2eV}, where the $L_\mathrm{2-10~ keV}$ and photon index $\Gamma$ are derived from X-ray spectra fitting. 
\begin{equation}
\alpha_\mathrm{ox} = \frac{\log (L_\mathrm{UVW1} / L_\mathrm{2keV} )} {\log (\nu_\mathrm{2keV} /  \nu_\mathrm{UVW1} )}=0.384\times {\log (L_\mathrm{UVW1} / L_\mathrm{2keV} )}
\label{definition_alpha_ox}
\end{equation}
\begin{eqnarray}
\large{
L_\mathrm{2~keV}= 
\begin{cases}\displaystyle\frac{L_\mathrm{2-10~ keV} (2-\Gamma)}{\nu_\mathrm{2~keV} \times (5^{2-\Gamma}-1)} \quad &, 
\Gamma \neq 2 \\ 
\displaystyle\frac{L_\mathrm{2-10~ keV}}{\nu_\mathrm{2~keV} \times \mathrm{ln} 5}\quad  &, \Gamma = 2
\end{cases} }
\label{definition_f2eV}
\end{eqnarray} 
The \alphaox\, and $\log{L_\mathrm{X}/L_\mathrm{Edd}}$ also follow a negative correlation in the type 1.9 phase, where the Spearman correlation coefficient is $-0.67$ ($p=1.7\times10^{-7}$). The data in type 1 phase and the re-flare also apparently deviate from the negative correlation (\autoref{fig:alpha_ox_lx}).
 

\subsection{Radio--X-ray luminosity correlation}
\begin{figure}
\centering
	\includegraphics[width=\linewidth]{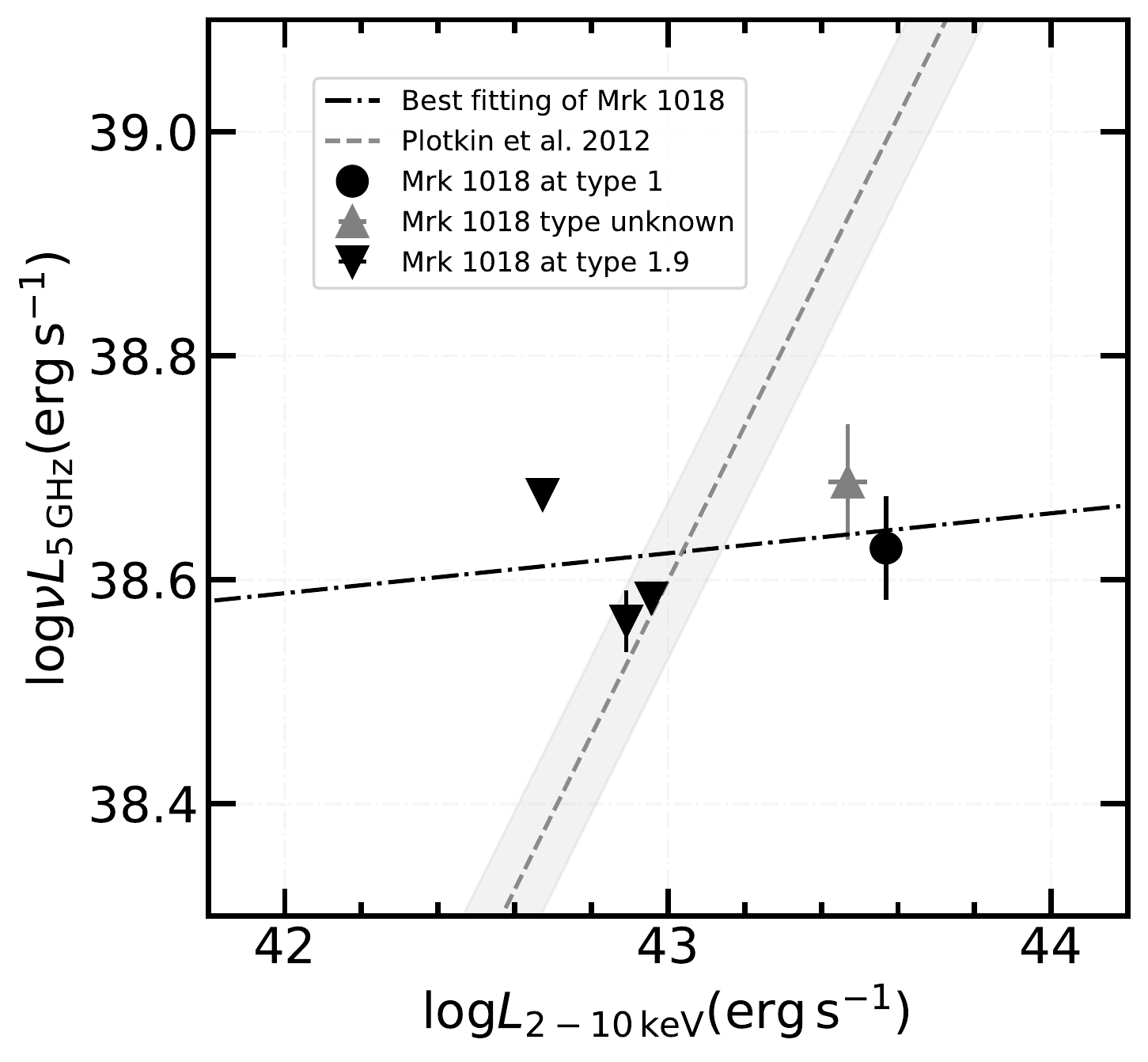}
    \caption{The $\log L_\mathrm{R}$-$\log L_\mathrm{X}$ correlation. The dash-dot line represents the best-fitting line of Mrk~1018 with a slope of $\sim 0.04$. 
    The fundamental plane of a sample of black holes in \citet{2012MNRAS.419..267P} with a slope of $\sim 0.6$ is presented in the grey dashed line with intrinsic $\sigma=0.07$ for comparison.} 
    \label{fig:radio-xray-mass_relation_Plotkin2012}
\end{figure}

The correlation of the 5 GHz radio luminosity ($\log L_\mathrm{R}$) and 2--10~keV X-ray luminosity ($\log L_\mathrm{X}$) is presented in \autoref{fig:radio-xray-mass_relation_Plotkin2012}, where the quasi-simultaneous radio and X-ray observations within 100 days are adopted. The radio and X--ray luminosity follow a quite flat correlation during the luminosity range of $L_\mathrm{X}/L_\mathrm{Edd}$ $\sim$ 5$\times 10^{-4}$-- 4$\times 10^{-3}$, where the Spearman correlation coefficient is $0.2$ ($p=0.75$). Coincidently, the two points in 2017 and average of the $\log L_\mathrm{R}$ and $\log L_\mathrm{X}$ correlation of Mrk 1018 roughly follows the fundamental plane defined by the sample of AGN and XRB \citep[e.g.][]{2012MNRAS.419..267P}. But the points before 2016 deviate the fundamental plane where $L_\mathrm{R}\propto L_\mathrm{X}^{0.6}$.

\section{Discussion}\label{sec:discussion}
\subsection{The spectral evolution and possible accretion mode transition }
\label{sec:spectral evolution}
The X-ray photon index can shed light on the physical properties of the hot plasma in the advection-dominated
accretion flow (ADAF) or corona above and below the cold disk, which is controlled mainly by the electron temperature and optical depth. The negative/positive correlation of $\Gamma$-$\log{L_\mathrm{X}/L_\mathrm{Edd}}$ below/above a critical value ($L_\mathrm{X}/L_\mathrm{Edd} \sim 10^{-3}$) has been reported in both AGNs and X-ray binaries (XRBs) \citep[e.g.][]{2008ApJ...682..212W,2009MNRAS.399..349G,2011A&A...530A.149Y,2015MNRAS.447.1692Y,2020ApJ...889L..18Y}. The reason for the opposite X-ray spectral behavior is thought to be the differences of the seed photons for the Compton scattering, i.e. the seed photons are from the synchrotron emission of the hot ADAF at the lower luminosity branch, while from the thermal emission from Shakura--Sunyaev disk \citep[SSD; e.g. ][]{2013ApJ...764....2Q} at the higher luminosity branch. There is an evident negative correlation of $\Gamma$-$\log{L_\mathrm{X}/L_\mathrm{Edd}}$ in the type 1.9 phase of Mrk~1018, which is similar to the low-luminosity AGNs and the low/hard state XRBs. The data in the type 1 phase deviate from the negative correlation evidently, where the physical origin for optical and X-ray emission should be changed. 

The optical/UV-to-X-ray spectral index \alphaox\, is a good indicator of the broadband spectral energy distribution (SED), where the optical/UV and X-ray emission come from the cold SSD (or outer truncated SSD) and hot corona/ADAF respectively in AGNs. The negative and positive correlations of \alphaox-$\log{L_\mathrm{X}/L_\mathrm{Edd}}$ have been found in low-luminosity AGNs \citep[e.g.][]{2011ApJ...739...64X,2017MNRAS.471.2848L} and luminous AGNs \citep[e.g.][]{2010A&A...512A..34L, 2013A&A...550A..71V,2016ApJ...819..154L}. The ADAF model can roughly explain the negative \alphaox\,--$L_\mathrm{bol}$ correlation at the lower luminosity branch \citep{2011ApJ...739...64X,2017MNRAS.471.2848L}, and the disk-corona model can explain the positive \alphaox\,--$L_\mathrm{bol}$ correlation at the higher luminosity branch \citep{2017A&A...602A..79L, 2018MNRAS.480.1247K,2019A&A...628A.135A}. \citet{2011MNRAS.413.2259S} simulated the spectral states of AGNs by analogy with BHXRBs and found that the simulated AGNs at different spectral states and luminosity roughly follow a ``V''-shape \alphaox--$L_\mathrm{bol}$ correlation \citep[negative/positive correlation below/above a critical value, see also in ][]{2019ApJ...883...76R}. In other words, the opposite correlations between \alphaox\, and $L$ also support the idea that the accretion mode changes in the high and low luminosity AGNs \citep[see][]{2011MNRAS.413.2259S,2019ApJ...883...76R}.
There is an evident negative correlation of \alphaox-$\log{L_\mathrm{X}/L_\mathrm{Edd}}$ in the type 1.9 AGN phase of Mrk~1018, which is similar to the low-luminosity AGNs. The data in type 1 phase apparently deviate from the negative correlation, which indicates that the accretion mode is different from type 1.9 phase. A much stronger SSD component in type 1 phase will lead to a higher \alphaox. 

The evolutions of both $\Gamma$-$\log{L_\mathrm{X}/L_\mathrm{Edd}}$ and \alphaox-$\log{L_\mathrm{X}/L_\mathrm{Edd}}$ suggest that the strong evolution of underlying accretion disk in Mrk 1018 during the type transition. The change of the broad emission lines is most possibly regulated by the appearance or disappearance of cold disk near the black hole horizon, which regulates the radiative efficiency of accretion flow and/or ionization luminosity for the clouds in broad line region.

\subsection{The variability timescale}
\label{sec:timescale}
In Mrk 1018, the broadband spectra are dramatically changed with the rapid decay of luminosity \citep[see also ][]{2016A&A...593L...9H,2018MNRAS.480.3898N} between 2010 and 2015.
More and more studies show that the spectral evolution of CL-AGNs somehow are quite similar to the spectral state transitions in Galactic BH XRBs \citep{2018MNRAS.480.3898N,2019arXiv190904676R,2019ApJ...883...76R,2020MNRAS.492.2335L}. The state transitions in BHXRBs normally occur on timescale of days to tens of days \citep{2009ApJ...701.1940Y,2010MNRAS.403...61D}, which is usually
attributed to the viscous timescale of the inner disk in the truncated accretion disk model \citep[see reviews in ][]{2007A&ARv..15....1D}. The viscous timescale $\tau_\mathrm{vis}$ for a $10^{8} M_{\odot}$ AGN is given by,
\begin{equation}
\tau_\mathrm{vis} \sim 5.7\times 10^{-3} \alpha^{-1}(\frac{M_\mathrm{BH}}{10^8M_{\odot}})(\frac{R_\mathrm{tr}}{R_g})^{3/2} (\frac{H}{R})^{-2} \, \mathrm{days} 
\end{equation} 
 The $H/R$ is the ratio between disk height and disk radius, the $R_\mathrm{tr}$ is the inner truncated radius. So if the truncated radius of the accretion disk is smaller than  $\sim$20$R_{g}$, the viscous timescale will roughly agree with the spectral evolution timescale ($\lesssim$5 years) of Mrk 1018 for $\alpha=0.1$ and $H/R=0.05$ \citep[e.g.][]{2020MNRAS.492.2335L}. Such a viscous timescale is too short for a standard disk, since the required $H/R$ is much larger than that expected in standard disk. So there are also some different scenarios proposed to account for such a short timescale, such as the thermal and heating/cooling front timescales in the innermost regions of the accretion disk \citep{2018ApJ...864...27S}, a radiation-pressure dominated accretion disk with faster sound speed and hence the viscous speed \citep{2018MNRAS.480.3898N}, a thick disk supported by the magnetic pressure \citep{2019MNRAS.483L..17D}, and the radiation pressure instability and transition between standard disk and ADAF \citep{2020A&A...641A.167S}. 
 
A re-flare with a timescale on a magnitude of hundreds of days during the decay phase was found, where the physical reason for the variability is unclear. The rise timescale ($\lesssim $ 100 days) of the re-flare is consistent with the thermal timescale at an emission distance of $R\lesssim 200R_g$.  A short ``turn-on'' (appearance of broad emission lines and increasing luminosity by a factor of 8) timescale $\sim$ 70 days of PS1-13cbe is also suggested to be compatible with the thermal timescale under the scenario of UV/X-ray reprocessing \citep[][]{2019MNRAS.487.4057K}. However, there is no information about the emission lines of Mrk 1018 during the re-flare, so we have no idea whether the AGN type has been changed during this re-flare. 

We estimated the time lag $\tau \sim 20$ days between UV and X-ray variations of Mrk 1018 in the type 1.9 phase, which is much longer than previous results in other nearby AGNs \citep[usually less than 5 days, e.g. ][]{2009MNRAS.397.2004A,2014ApJ...788...48S,2017MNRAS.464.3194B,2017ApJ...840...41E,2019ApJ...870..123E,2020MNRAS.494.1165L}. The X-ray reprocessing is usually thought to be the main mechanism for the optical/UV variation \citep[e.g.][]{1991ApJ...371..541K,2007MNRAS.380..669C,2021MNRAS.503.4163K}, where the optical/UV flux is expected to vary in response to X-ray after a light-crossing time. In the X-ray reprocessing scenario \citep{2007MNRAS.380..669C}, the time delay is expected to be positively correlated with the wavelength, which is not found in Mrk~1018 (see \autoref{tab:tablelag}). The time lag $\sim 20$ days corresponds to a light-crossing size $\sim$ 5000 $R_g$ for Mrk 1018, which is also much larger than the typical disk size of AGNs. Such a long UV/X-ray time lag is approximately consistent with the thermal timescale at $R\sim 65 R_g$ for Mrk 1018, which may suggest a slower physical process, such as the thermal fluctuation or instability \citep{2009ApJ...698..895K,2017MNRAS.470.3591G}.

\subsection{Radio--X-ray correlation}
The simultaneous X-ray and radio emission track the connection between accretion and ejection activities of CL-AGNs \citep[e.g.][]{2016MNRAS.460..304K,2021MNRAS.tmp..700Y}.
It is widely accepted that there is a non-linear correlation between the 5 GHz radio luminosity and the 2--10 keV X-ray luminosity spanning from the super-massive black holes to stellar mass black holes: $L_\mathrm{R}\propto L_\mathrm{X}^{0.6}$ \citep[e.g.][]{2003MNRAS.345.1057M,2004A&A...414..895F}. There are also some sources following a steeper radio--X-ray correlation with a power-law index of $\sim$1.4 at higher luminosity, a flat radio--X-ray correlation with a power-law index of $\sim$0 at transition stage \citep[e.g.][]{2011MNRAS.414..677C,2014ApJ...788...52C,2016MNRAS.463.2287X}. The ``hybrid'' radio--X-ray correlation is possibly regulated by the underlying accretion processes \citep[e.g.][]{2016MNRAS.456.4377X} or due to different jet properties \citep[e.g.][]{2018MNRAS.481.4513I,2018MNRAS.473.4122E}. It should be noted that Mrk~1018 roughly follows the radio--X-ray correlation as defined by other BH sources \citep{2012MNRAS.419..267P} even though its own correlation is quite flat. The X-ray luminosity's decline by a factor $\sim $7.5 requires a variability by a factor of $\sim$ 3.3 in radio luminosity to follow the radio--X-ray correlation with a slope of $0.6$, which is far beyond the radio variability of Mrk~1018. The possible physical reasons include: 1) the radio variability timescale is much longer than X-ray since that the radio emission comes from a larger scale of jet; 2) the flat radio--X-ray correlation is mainly regulated by the X-ray emission, where the accretion rate does not vary much but the radiative efficiency change a lot when the accretion rate is close to a critical value. The Eddington ratio of bolometric luminosity is around 1\% \citep[][]{2018MNRAS.480.3898N}, which suggests that the radiative efficiency in Mrk~1018 may indeed easily suffer strong variations due to the transition of accretion modes.

\section{Summary}
\label{sec:conclusion}
The main results are summarized as follows,
\begin{enumerate}

\item We present the long-term and multi-wavelength variability from radio to X-ray band for a CL-AGN of Mrk 1018. We find a re-flare in both optical/UV and X-ray bands during the decay phase and a time lag $\sim$ 20 days of optical/UV behind X-ray variations during the type 1.9 phase.    

\item We find negative correlations of $\Gamma$-$L_\mathrm{X}/L_\mathrm{Edd}$ and \alphaox-$L_\mathrm{X}/L_\mathrm{Edd}$ in the type 1.9 phase, which are consistent with the prediction of the radiatively inefficient accretion (e.g., ADAF). The data in the type 1 phase deviate from the negative correlations, where the accretion mode may change into a radiatively efficient disk-corona system. Therefore, the change of broad emission lines might be regulated by the underlying accretion process. 

\item The radio emission in CL-AGN of Mrk~1018 was roughly unchanged before the type 1.9 phase (2005--2015), then it slightly declined about $\sim$ 20\% during 2016--2017. Therefore, radio--X-ray correlation is quite flat in Mrk 1018, which is much shallower than that found in low-luminosity AGNs and low-hard state XRBs with a slope of $\sim 0.6$. 

\end{enumerate}

\section*{Acknowledgements}
We thank the anonymous referee for useful comments and suggestions. We thank Prof Minfeng Gu for the discussions on the radio observations of CL-AGNs and Dr Linhui Wu and Minhua Zhou for discussions on the VLA data reduction. B.L. and Q.W. are supported in part by the NSFC (grant U1931203); Z.Y. is supported in part by the Natural Science Foundation of China (grants 11773055, U1938114), the Youth Innovation Promotion Association of CAS (ids. 2020265); W.Y. would like to acknowledge the support in part by the National Program on Key Research and Development Project (Grant No.2016YFA0400804) and the National Natural Science Foundation of China (grant number 11333005 and U1838203).


\section*{Data Availability}
The data underlying this article are available through HEASARC Browse database and NRAO Science Data Archive.


\input{ref.bbl}
\bibliographystyle{mnras}




\bsp	
\label{lastpage}
\appendix
\section{Appendix}

\begin{center}
\footnotesize
\onecolumn
\begin{longtable}{cccccc}
\caption{{\bfseries X-ray data of Mrk~1018. } Columns include the date of observations, observation ID, the telescope, the photon index $\Gamma$ with 90\% uncertainty, and Galactic-absorption corrected 2--10~keV X-ray flux.
\label{tab:tablexray}} \\
\hline 
\hline
 Date   &   ObsID & Telescope   &  $\Gamma$ & $F_{\rm{2-10~keV}}$ &    \\ 
 (MJD)  &         &             &           & [$10^{-12}$ erg cm$^{-2}$\rm{s}$^{-1}$] &  \\ 
\hline
\endfirsthead
\multicolumn{5}{c}%
{{\bfseries \tablename\ \thetable{} -- continued from previous page}} \\
\hline 
\hline
 Date   &   ObsID & Telescope   &  $\Gamma$ & $F_{\rm{2-10~keV}}$ &    \\ 
 (MJD)  &         &             &           & [$10^{-12}$ erg cm$^{-2}$\rm{s}$^{-1}$] &  \\ 
\hline
\endhead
\hline 
\multicolumn{5}{c}{
The superscripts $^{(a)}$ and $^{(b)}$ represent the fitting results from \citet{2012ApJ...745..107W} and \citet{2016A&A...593L...9H}, repectively.}\\ \
\endfoot
\hline 
\multicolumn{5}{c}{Notes: The facility is represented by C-\chandra, S-\swift, X-\xmm, N-\nustar and Su-\suzaku.}\\ \
\endlastfoot

53385 & 201090201 & X & 1.68 $\pm$ 0.11 & 10.40 $\pm$ 0.50 & \\
53587 & 35166001 & S & 1.94 $\pm$ 0.06 & 11.75 $\pm$ 0.81 & \\
54273 & 30955002 & S & 1.91 $\pm$ 0.05 & 8.91 $\pm$ 0.62 & \\
54275 & 30955003 & S & 1.95 $\pm$ 0.04 & 8.91 $\pm$ 0.41 & \\
54628 & 35776001 & S & 1.73 $\pm$ 0.04 & 10.72 $\pm$ 0.49 & \\
54685 & 554920301 & X & 1.79 $\pm$ 0.03 & 11.50 $\pm$ 0.20 & \\
55015 & 704044010$^{a}$ & Su  & 2.00 $\pm$ 0.03 & 10.00 $\pm$ 0.50  \\
55527 & 12868$^{b}$ & C  & 1.68 $\pm$ 0.04 & 9.20 $\pm$ 0.20  \\
56352 & 49654001 & S & 1.82 $\pm$ 0.23 & 2.51 $\pm$ 0.75 & \\
56450 & 49654002 & S & 1.45 $\pm$ 0.09 & 7.94 $\pm$ 0.91 & \\
56817 & 49654004 & S & 1.39 $\pm$ 0.19 & 1.86 $\pm$ 0.39 & \\
57428 & 60160087002 & N & 1.85 $\pm$ 0.08 & 1.80 $\pm$ 0.10 & \\
57429 & 80898001 & S & 1.72 $\pm$ 0.12 & 1.62 $\pm$ 0.26 & \\
57434 & 80898002 & S & 1.46 $\pm$ 0.12 & 2.19 $\pm$ 0.35 & \\
57443 & 18789 & C & 1.68 $\pm$ 0.03 & 1.27 $\pm$ 0.03 & \\
57801 & 19560 & C & 1.61 $\pm$ 0.02 & 2.44 $\pm$ 0.02 & \\
58123 & 60301022002 & N & 1.80 $\pm$ 0.06 & 2.10 $\pm$ 0.10 & \\
58124 & 88207001 & S & 1.63 $\pm$ 0.15 & 2.04 $\pm$ 0.38 & \\
58126 & 20366 & C & 1.60 $\pm$ 0.03 & 1.79 $\pm$ 0.04 & \\
58173 & 88207002 & S & 1.60 $\pm$ 0.16 & 2.00 $\pm$ 0.41 & \\
58180 & 20367 & C & 1.63 $\pm$ 0.04 & 1.56 $\pm$ 0.03 & \\
58182 & 60301022003 & N & 1.80 $\pm$ 0.06 & 1.50 $\pm$ 0.07 & \\
58281 & 20368 & C & 1.64 $\pm$ 0.03 & 1.74 $\pm$ 0.04 & \\
58316 & 60301022005 & N & 1.74 $\pm$ 0.05 & 2.00 $\pm$ 0.09 & \\
58316 & 88207003 & S & 1.63 $\pm$ 0.14 & 2.19 $\pm$ 0.40 & \\
58362 & 35776002 & S & 1.98 $\pm$ 0.29 & 1.00 $\pm$ 0.37 & \\
58363 & 35776003 & S & 1.41 $\pm$ 0.29 & 1.74 $\pm$ 0.64 & \\
58364 & 35776004 & S & 1.34 $\pm$ 0.23 & 2.40 $\pm$ 0.66 & \\
58365 & 35776005 & S & 1.48 $\pm$ 0.23 & 1.82 $\pm$ 0.50 & \\
58368 & 35776006 & S & 1.44 $\pm$ 0.18 & 2.63 $\pm$ 0.61 & \\
58369 & 35776007 & S & 1.64 $\pm$ 0.19 & 2.63 $\pm$ 0.61 & \\
58370 & 20369 & C & 1.63 $\pm$ 0.03 & 2.65 $\pm$ 0.06 & \\
58370 & 35776008 & S & 1.81 $\pm$ 0.25 & 2.14 $\pm$ 0.69 & \\
58374 & 35776010 & S & 1.56 $\pm$ 0.34 & 1.86 $\pm$ 0.81 & \\
58375 & 35776011 & S & 1.39 $\pm$ 0.23 & 2.04 $\pm$ 0.56 & \\
58378 & 35776014 & S & 1.33 $\pm$ 0.21 & 2.95 $\pm$ 0.75 & \\
58384 & 35776015 & S & 1.60 $\pm$ 0.35 & 2.82 $\pm$ 1.30 & \\
58385 & 35776016 & S & 1.16 $\pm$ 0.21 & 3.63 $\pm$ 0.92 & \\
58390 & 35776017 & S & 1.49 $\pm$ 0.24 & 1.86 $\pm$ 0.56 & \\
58390 & 35776018 & S & 1.72 $\pm$ 0.22 & 1.74 $\pm$ 0.48 & \\
58391 & 35776019 & S & 1.51 $\pm$ 0.29 & 1.51 $\pm$ 0.52 & \\
58392 & 35776020 & S & 1.50 $\pm$ 0.27 & 1.48 $\pm$ 0.51 & \\
58393 & 35776021 & S & 1.13 $\pm$ 0.26 & 2.57 $\pm$ 0.77 & \\
58395 & 35776023 & S & 1.74 $\pm$ 0.27 & 1.48 $\pm$ 0.51 & \\
58396 & 35776024 & S & 1.82 $\pm$ 0.27 & 1.38 $\pm$ 0.48 & \\
58399 & 35776026 & S & 1.64 $\pm$ 0.24 & 2.14 $\pm$ 0.64 & \\
58399 & 35776027 & S & 1.97 $\pm$ 0.27 & 1.20 $\pm$ 0.42 & \\
58402 & 35776029 & S & 1.74 $\pm$ 0.27 & 1.70 $\pm$ 0.63 & \\
58409 & 35776032 & S & 1.02 $\pm$ 0.29 & 2.09 $\pm$ 0.72 & \\
58410 & 35776033 & S & 1.98 $\pm$ 0.28 & 0.87 $\pm$ 0.32 & \\
58412 & 35776034 & S & 1.01 $\pm$ 0.24 & 3.24 $\pm$ 0.89 & \\
58413 & 35776035 & S & 1.36 $\pm$ 0.26 & 1.82 $\pm$ 0.59 & \\
58417 & 35776036 & S & 1.12 $\pm$ 0.27 & 2.14 $\pm$ 0.69 & \\
58418 & 35776037 & S & 2.03 $\pm$ 0.42 & 0.63 $\pm$ 0.31 & \\
58420 & 35776038 & S & 1.72 $\pm$ 0.44 & 0.91 $\pm$ 0.50 & \\
58420 & 35776039 & S & 1.33 $\pm$ 0.32 & 1.51 $\pm$ 0.56 & \\
58422 & 35776040 & S & 1.30 $\pm$ 0.28 & 1.55 $\pm$ 0.53 & \\
58423 & 35776041 & S & 1.55 $\pm$ 0.32 & 1.41 $\pm$ 0.59 & \\
58424 & 35776042 & S & 1.94 $\pm$ 0.22 & 1.17 $\pm$ 0.35 & \\
58425 & 35776043 & S & 1.78 $\pm$ 0.31 & 0.98 $\pm$ 0.38 & \\
58426 & 35776044 & S & 1.15 $\pm$ 0.35 & 2.00 $\pm$ 0.87 & \\
58427 & 35776045 & S & 1.51 $\pm$ 0.33 & 1.10 $\pm$ 0.45 & \\
58428 & 35776046 & S & 1.48 $\pm$ 0.29 & 1.51 $\pm$ 0.52 & \\
58429 & 35776047 & S & 1.38 $\pm$ 0.30 & 1.74 $\pm$ 0.60 & \\
58430 & 20370 & C & 1.64 $\pm$ 0.04 & 1.39 $\pm$ 0.03 & \\
58433 & 35776049 & S & 1.76 $\pm$ 0.38 & 1.10 $\pm$ 0.50 & \\
58434 & 35776050 & S & 1.56 $\pm$ 0.30 & 1.23 $\pm$ 0.45 & \\
58436 & 35776051 & S & 1.57 $\pm$ 0.26 & 1.48 $\pm$ 0.48 & \\
58437 & 35776052 & S & 1.17 $\pm$ 0.30 & 1.82 $\pm$ 0.63 & \\
58439 & 35776054 & S & 1.65 $\pm$ 0.29 & 1.48 $\pm$ 0.51 & \\
58444 & 35776056 & S & 1.21 $\pm$ 0.25 & 2.51 $\pm$ 0.75 & \\
58445 & 35776057 & S & 1.99 $\pm$ 0.42 & 0.68 $\pm$ 0.34 & \\
58446 & 35776058 & S & 1.27 $\pm$ 0.43 & 1.70 $\pm$ 0.98 & \\
58520 & 21432 & C & 1.66 $\pm$ 0.03 & 1.71 $\pm$ 0.03 & \\
58521 & 22082 & C & 1.65 $\pm$ 0.03 & 1.87 $\pm$ 0.04 & \\
58663 & 35776059 & S & 1.69 $\pm$ 0.28 & 0.79 $\pm$ 0.27 & \\
\hline

\end{longtable}

\end{center}

\begin{center}
\centering
\footnotesize
\onecolumn
\begin{longtable}{ccccccccc}
\caption{{ \bf UVOT data of Mrk~1018. } Columns include the date of observations and the absorption and host corrected flux. 
\label{tab:tableuvot}} \\
\hline
\hline
Date   &  U     & Date     &  UVW1  &  Date  &  UVM2   &  Date  &   UVW2 &   \\ 
(MJD)  &  [mJy] &  (MJD)   &[mJy]   & (MJD)  & [mJy]   & (MJD)  & [mJy]  & \\ 
\hline
\endfirsthead
\multicolumn{8}{c}%
{{\bfseries \tablename\ \thetable{} -- continued from previous page}} \\
\hline 
\hline
Date   &  U     & Date     &  UVW1  &  Date  &  UVM2   &  Date  & UVW2   &   \\ 
(MJD)  &  [mJy] &  (MJD)   &[mJy]   & (MJD)  & [mJy]   & (MJD)  & [mJy]  & \\ \\ \hline 
\endhead
53587 & 3.77 $\pm$ 0.10 & 53587 & 3.56 $\pm$ 0.11 & 53587 & 3.51 $\pm$ 0.06 & 53587 & 3.53 $\pm$ 0.08 & \\
54628 & 2.86 $\pm$ 0.08 & 54271 & 2.99 $\pm$ 0.09 & 54628 & 2.46 $\pm$ 0.04 & 54273 & 2.92 $\pm$ 0.06 & \\
56352 & 0.35 $\pm$ 0.04 & 54275 & 3.12 $\pm$ 0.09 & 56352 & 0.34 $\pm$ 0.02 & 54628 & 2.35 $\pm$ 0.05 & \\
56450 & 0.47 $\pm$ 0.03 & 54628 & 2.60 $\pm$ 0.08 & 56450 & 0.41 $\pm$ 0.02 & 56352 & 0.24 $\pm$ 0.01 & \\
56817 & 0.08 $\pm$ 0.02 & 56352 & 0.32 $\pm$ 0.02 & 56817 & 0.09 $\pm$ 0.01 & 56450 & 0.39 $\pm$ 0.01 & \\
57429 & 0.12 $\pm$ 0.02 & 56450 & 0.48 $\pm$ 0.02 & 57429 & 0.07 $\pm$ 0.01 & 56817 & 0.09 $\pm$ 0.01 & \\
57434 & 0.05 $\pm$ 0.02 & 56817 & 0.13 $\pm$ 0.01 & 57434 & 0.06 $\pm$ 0.00 & 57429 & 0.06 $\pm$ 0.00 & \\
58124 & 0.07 $\pm$ 0.01 & 57429 & 0.09 $\pm$ 0.01 & 58362 & 0.07 $\pm$ 0.01 & 57434 & 0.06 $\pm$ 0.01 & \\
58316 & 0.07 $\pm$ 0.01 & 57434 & 0.08 $\pm$ 0.01 & 58363 & 0.06 $\pm$ 0.01 & 58173 & 0.08 $\pm$ 0.00 & \\
58362 & 0.12 $\pm$ 0.03 & 58362 & 0.12 $\pm$ 0.02 & 58364 & 0.07 $\pm$ 0.01 & 58362 & 0.08 $\pm$ 0.01 & \\
58363 & 0.05 $\pm$ 0.02 & 58363 & 0.09 $\pm$ 0.01 & 58365 & 0.08 $\pm$ 0.01 & 58363 & 0.08 $\pm$ 0.01 & \\
58364 & 0.09 $\pm$ 0.03 & 58364 & 0.10 $\pm$ 0.01 & 58368 & 0.06 $\pm$ 0.01 & 58364 & 0.07 $\pm$ 0.01 & \\
58365 & 0.03 $\pm$ 0.02 & 58365 & 0.10 $\pm$ 0.01 & 58369 & 0.05 $\pm$ 0.01 & 58365 & 0.08 $\pm$ 0.01 & \\
58368 & 0.07 $\pm$ 0.02 & 58368 & 0.08 $\pm$ 0.01 & 58370 & 0.06 $\pm$ 0.01 & 58368 & 0.07 $\pm$ 0.01 & \\
58369 & 0.06 $\pm$ 0.02 & 58369 & 0.09 $\pm$ 0.01 & 58374 & 0.06 $\pm$ 0.01 & 58369 & 0.05 $\pm$ 0.01 & \\
58370 & 0.05 $\pm$ 0.03 & 58370 & 0.13 $\pm$ 0.02 & 58375 & 0.05 $\pm$ 0.01 & 58370 & 0.08 $\pm$ 0.01 & \\
58374 & 0.09 $\pm$ 0.03 & 58374 & 0.10 $\pm$ 0.01 & 58375 & 0.06 $\pm$ 0.01 & 58374 & 0.07 $\pm$ 0.01 & \\
58375 & 0.08 $\pm$ 0.03 & 58375 & 0.07 $\pm$ 0.02 & 58378 & 0.07 $\pm$ 0.01 & 58375 & 0.07 $\pm$ 0.01 & \\
58375 & 0.10 $\pm$ 0.03 & 58375 & 0.09 $\pm$ 0.01 & 58384 & 0.07 $\pm$ 0.01 & 58375 & 0.09 $\pm$ 0.01 & \\
58378 & 0.04 $\pm$ 0.02 & 58378 & 0.09 $\pm$ 0.01 & 58385 & 0.09 $\pm$ 0.01 & 58378 & 0.07 $\pm$ 0.01 & \\
58384 & 0.07 $\pm$ 0.03 & 58384 & 0.09 $\pm$ 0.01 & 58390 & 0.06 $\pm$ 0.01 & 58384 & 0.07 $\pm$ 0.01 & \\
58385 & 0.04 $\pm$ 0.03 & 58385 & 0.09 $\pm$ 0.02 & 58390 & 0.07 $\pm$ 0.01 & 58385 & 0.07 $\pm$ 0.01 & \\
58390 & 0.08 $\pm$ 0.02 & 58390 & 0.14 $\pm$ 0.02 & 58391 & 0.07 $\pm$ 0.01 & 58390 & 0.08 $\pm$ 0.01 & \\
58390 & 0.07 $\pm$ 0.02 & 58390 & 0.12 $\pm$ 0.02 & 58392 & 0.06 $\pm$ 0.01 & 58390 & 0.07 $\pm$ 0.01 & \\
58391 & 0.08 $\pm$ 0.03 & 58391 & 0.14 $\pm$ 0.02 & 58393 & 0.09 $\pm$ 0.01 & 58391 & 0.08 $\pm$ 0.01 & \\
58392 & 0.07 $\pm$ 0.02 & 58392 & 0.13 $\pm$ 0.02 & 58395 & 0.06 $\pm$ 0.01 & 58392 & 0.08 $\pm$ 0.01 & \\
58393 & 0.12 $\pm$ 0.03 & 58393 & 0.14 $\pm$ 0.02 & 58396 & 0.06 $\pm$ 0.01 & 58393 & 0.07 $\pm$ 0.01 & \\
58395 & 0.12 $\pm$ 0.03 & 58395 & 0.10 $\pm$ 0.01 & 58397 & 0.06 $\pm$ 0.01 & 58395 & 0.08 $\pm$ 0.01 & \\
58396 & 0.05 $\pm$ 0.03 & 58396 & 0.11 $\pm$ 0.02 & 58399 & 0.07 $\pm$ 0.01 & 58396 & 0.06 $\pm$ 0.01 & \\
58397 & 0.10 $\pm$ 0.03 & 58397 & 0.11 $\pm$ 0.01 & 58399 & 0.10 $\pm$ 0.01 & 58397 & 0.06 $\pm$ 0.01 & \\
58399 & 0.09 $\pm$ 0.03 & 58399 & 0.12 $\pm$ 0.02 & 58402 & 0.08 $\pm$ 0.01 & 58399 & 0.09 $\pm$ 0.01 & \\
58399 & 0.07 $\pm$ 0.02 & 58399 & 0.10 $\pm$ 0.01 & 58406 & 0.06 $\pm$ 0.01 & 58399 & 0.08 $\pm$ 0.01 & \\
58402 & 0.04 $\pm$ 0.02 & 58402 & 0.11 $\pm$ 0.01 & 58409 & 0.09 $\pm$ 0.01 & 58402 & 0.07 $\pm$ 0.01 & \\
58406 & 0.03 $\pm$ 0.02 & 58406 & 0.11 $\pm$ 0.02 & 58410 & 0.07 $\pm$ 0.01 & 58406 & 0.08 $\pm$ 0.01 & \\
58409 & 0.00 $\pm$ 0.02 & 58409 & 0.09 $\pm$ 0.01 & 58412 & 0.05 $\pm$ 0.01 & 58409 & 0.07 $\pm$ 0.01 & \\
58410 & 0.07 $\pm$ 0.02 & 58410 & 0.10 $\pm$ 0.01 & 58413 & 0.05 $\pm$ 0.01 & 58410 & 0.08 $\pm$ 0.01 & \\
58412 & 0.07 $\pm$ 0.02 & 58412 & 0.09 $\pm$ 0.01 & 58417 & 0.05 $\pm$ 0.01 & 58412 & 0.06 $\pm$ 0.01 & \\
58413 & 0.06 $\pm$ 0.02 & 58413 & 0.10 $\pm$ 0.01 & 58418 & 0.06 $\pm$ 0.01 & 58413 & 0.07 $\pm$ 0.01 & \\
58417 & 0.09 $\pm$ 0.03 & 58417 & 0.07 $\pm$ 0.01 & 58420 & 0.07 $\pm$ 0.01 & 58417 & 0.06 $\pm$ 0.01 & \\
58418 & 0.06 $\pm$ 0.02 & 58418 & 0.09 $\pm$ 0.01 & 58420 & 0.06 $\pm$ 0.01 & 58418 & 0.07 $\pm$ 0.01 & \\
58420 & 0.01 $\pm$ 0.02 & 58420 & 0.08 $\pm$ 0.01 & 58422 & 0.07 $\pm$ 0.01 & 58420 & 0.06 $\pm$ 0.01 & \\
58420 & 0.07 $\pm$ 0.02 & 58420 & 0.07 $\pm$ 0.01 & 58423 & 0.06 $\pm$ 0.01 & 58420 & 0.06 $\pm$ 0.01 & \\
58422 & 0.02 $\pm$ 0.02 & 58422 & 0.07 $\pm$ 0.01 & 58424 & 0.07 $\pm$ 0.01 & 58422 & 0.06 $\pm$ 0.01 & \\
58423 & 0.10 $\pm$ 0.03 & 58423 & 0.07 $\pm$ 0.01 & 58425 & 0.07 $\pm$ 0.01 & 58423 & 0.07 $\pm$ 0.01 & \\
58424 & 0.06 $\pm$ 0.02 & 58424 & 0.08 $\pm$ 0.01 & 58426 & 0.05 $\pm$ 0.01 & 58424 & 0.06 $\pm$ 0.01 & \\
58425 & 0.03 $\pm$ 0.02 & 58425 & 0.09 $\pm$ 0.01 & 58427 & 0.06 $\pm$ 0.01 & 58425 & 0.08 $\pm$ 0.01 & \\
58426 & 0.05 $\pm$ 0.02 & 58426 & 0.07 $\pm$ 0.01 & 58428 & 0.06 $\pm$ 0.01 & 58426 & 0.06 $\pm$ 0.01 & \\
58427 & 0.06 $\pm$ 0.02 & 58427 & 0.12 $\pm$ 0.01 & 58429 & 0.05 $\pm$ 0.01 & 58427 & 0.07 $\pm$ 0.01 & \\
58428 & 0.06 $\pm$ 0.02 & 58428 & 0.09 $\pm$ 0.01 & 58433 & 0.08 $\pm$ 0.01 & 58428 & 0.06 $\pm$ 0.01 & \\
58429 & 0.05 $\pm$ 0.02 & 58429 & 0.10 $\pm$ 0.01 & 58434 & 0.08 $\pm$ 0.01 & 58429 & 0.07 $\pm$ 0.01 & \\
58433 & 0.06 $\pm$ 0.03 & 58433 & 0.10 $\pm$ 0.02 & 58436 & 0.07 $\pm$ 0.01 & 58433 & 0.07 $\pm$ 0.01 & \\
58434 & 0.03 $\pm$ 0.02 & 58434 & 0.09 $\pm$ 0.01 & 58437 & 0.07 $\pm$ 0.01 & 58434 & 0.08 $\pm$ 0.01 & \\
58436 & 0.06 $\pm$ 0.02 & 58436 & 0.08 $\pm$ 0.01 & 58439 & 0.06 $\pm$ 0.01 & 58436 & 0.07 $\pm$ 0.01 & \\
58437 & 0.06 $\pm$ 0.03 & 58437 & 0.10 $\pm$ 0.01 & 58441 & 0.06 $\pm$ 0.01 & 58437 & 0.06 $\pm$ 0.01 & \\
58439 & 0.09 $\pm$ 0.02 & 58439 & 0.11 $\pm$ 0.01 & 58444 & 0.04 $\pm$ 0.01 & 58439 & 0.07 $\pm$ 0.01 & \\
58441 & 0.04 $\pm$ 0.03 & 58441 & 0.09 $\pm$ 0.02 & 58445 & 0.06 $\pm$ 0.01 & 58441 & 0.08 $\pm$ 0.01 & \\
58444 & 0.04 $\pm$ 0.02 & 58444 & 0.10 $\pm$ 0.01 & 58446 & 0.05 $\pm$ 0.01 & 58444 & 0.07 $\pm$ 0.01 & \\
58445 & 0.08 $\pm$ 0.03 & 58445 & 0.08 $\pm$ 0.01 & 58663 & 0.03 $\pm$ 0.01 & 58445 & 0.06 $\pm$ 0.01 & \\
58446 & 0.10 $\pm$ 0.03 & 58446 & 0.07 $\pm$ 0.01 & & & 58446 & 0.04 $\pm$ 0.01 & \\
58663 & 0.05 $\pm$ 0.02 & 58663 & 0.07 $\pm$ 0.01 & & & 58663 & 0.04 $\pm$ 0.01 & \\

\hline
\end{longtable}
\end{center}

\begin{table*}
\centering
\caption{{\bf VLA data of Mrk~1018.} Columns include the date of observation, project name, band, frequency, integrated flux, radio spectral index ($\alpha_\mathrm{R}$) and references.}
\label{tab:tableradio}
\begin{tabular}{lcccccr}
\hline
\hline
 Date &  project & band  & Frequency  &$F_{\rm int}$   & $\alpha_\mathrm{R}$ & References  \\ 
 (MJD)&         &  &   [GHz]   &  [mJy]     &                 &         \\ \hline
\multirow{2}*{46032} & \multirow{2}*{AU0020}& L & 1.49 & 4.21 $\pm$ 0.23 & \multirow{2}*{0.52 $\pm0.07$} & \\
      &        & C & 4.86 & 2.29 $\pm$ 0.14 &  & \\ \hline
47261 & AB0476 & C & 4.86 & 1.91 $\pm$ 0.23 &  & \\
47692 & AB0540A & C & 4.86 & 2.62 $\pm$ 0.16 &  & \\
47732 & AB0540B & C & 4.86 & 2.31 $\pm$ 0.17 &  & \\
49341 & AC0308 & L & 1.4 & 4.20 $\pm$ 0.54 &  &  \citet{2002AJ....124..675C} \\
50031 & AB0628 & L & 1.4 & 4.20 $\pm$ 0.45 &  & \citet{1998AJ....115.1693C}\\
50970 & AB0878 & X & 8.46 & 2.47 $\pm$ 0.17 & 0.30 $\pm$ 0.08 & \\
52490 & AB0950 & L & 1.4 & 4.15 $\pm$ 0.25 &  &  \citet{2003yCat.8071....0B}\\
54873 & AR685 & L & 1.4 & 3.69 $\pm$ 0.19 &  &  \citet{2011AJ....142....3H}\\
54926 & AB1314 & L & 1.4 & 3.36 $\pm$ 0.20 &  &  \citet{2012yCat.8090....0B}\\
56542 & 13B-272 & L & 1.4 & 3.85 $\pm$ 0.31 &  &  \citet{2016MNRAS.460.4433H}\\\hline

\multirow{2}*{57481}&  \multirow{2}*{16A-444} & C & 5.0 & 2.56 $\pm$ 0.13 & \multirow{2}*{0.25 $\pm$ 0.10} & \\
      &         & X & 10.0 & 2.16 $\pm$ 0.11 &  & \\\hline
57719 & 16B-084 & X & 10.0 & 1.78 $\pm$ 0.09 &  & \\
57731 & 16B-084 & X & 10.0 & 1.97 $\pm$ 0.10 &  & \\
57768 & 16B-084 & C & 5.0 & 2.07 $\pm$ 0.10 &  0 & \\
58087 & VLASS1.1 & L & 3.0 & 2.30 $\pm$ 0.36 &  & \\

\hline 
\end{tabular}   
\end{table*}

\begin{table*}
\centering
\caption{{\bf Radio and X-ray luminosity correlation diagram.} Columns include the date of radio observation, rescaled radio flux at 5$\,$GHz ($F_{\rm{5GHz}}$)  , the date of X-ray observation,  X-ray flux in 2-10~keV band, the observation interval between two bands, the radio luminosity rescaled to 5 GHz ($L_R=\nu L_{\rm{5GHz}}$) and X-ray luminosity in 2-10~keV band ($L_\mathrm{X}$).}
\label{tab:radio_xray}
\begin{tabular}{lcccccc}
\hline
\hline

$T_{\rm Radio}$ & $F_{\rm 5GHz}$  & $T_{\rm X-ray}$  & $F_{2-10\rm keV}$ & $\delta$ T & log($L_{R}$) &log($L_{X}$) \\ 
(MJD) & [mJy]& (MJD) &[$10^{-12}$ erg cm$^{-2}$\rm{s}$^{-1}$]   &(Day)   & [erg$~s^{-1}$] &[erg$~s^{-1}$] \\
\hline

54926 & 2.29 $\pm$ 0.24 & 55015 & 10.00 $\pm$ 0.50 & -89 & 38.63 & 43.57 \\
56542 & 2.63 $\pm$ 0.31 & 56450 & 7.94 $\pm$ 0.91 & 91 & 38.69 & 43.47 \\
57481 & 2.56 $\pm$ 0.01 & 57443 & 1.27 $\pm$ 0.03 & 38 & 38.68 & 42.67 \\
57768 & 2.07 $\pm$ 0.01 & 57801 & 2.44 $\pm$ 0.02 & -33 & 38.58 & 42.96 \\
58087 & 1.97 $\pm$ 0.12 & 58123 & 2.10 $\pm$ 0.10 & -36 & 38.56 & 42.89 \\

 \hline
\end{tabular}\\
\end{table*}

\end{document}